\documentclass[12pt]{article}
\usepackage{graphicx}

\makeatletter

\def\amu{a_\mu}
\def\amuh{a_\mu^{\mathrm{had}}}

\def\aemz{\alpha_\mathrm{em}(M_Z)}

\def\dah{\Delta\alpha^{(5)}_{\rm had}}

\def\dah0{\Delta\alpha^{(5)}_{\rm had}(-s_0)}

\def\sfs{spectral functions}

\newcommand{\dmrho}{m_{\rho^\pm}-m_{\rho^0}}
\newcommand{\dgrho}{\Gamma_{\rho^\pm}-\Gamma_{\rho^0}}

\newcommand{\dalh}{\Delta \alpha^{\rm had}}

\newcommand{\mv}{\mbox{MeV}}

\newcommand{\epm}{e^+e^-}

\newcommand{\be}{\begin{equation}}
\newcommand{\ee}{\end{equation}}
\newcommand{\ba}{\begin{eqnarray}}
\newcommand{\ea}{\end{eqnarray}}
\newcommand{\bea}{\begin{eqnarray*}}
\newcommand{\eea}{\end{eqnarray*}}
\newcommand{\bet}{\begin{center} \begin{tabular}}
\newcommand{\ent}{\end{tabular} \end{center}}
\newcommand{\bb}{\begin{thebibliography}}
\newcommand{\eb}{\end{thebibliography}}

\newcommand{\bit}{\begin{itemize}}
\newcommand{\eit}{\end{itemize}}

\newcommand{\ha}{\frac{1}{2}}

\newcommand{\bary}{\begin{array}}
\newcommand{\eary}{\end{array}}

\renewcommand{\thefootnote}{\fnsymbol{footnote}}
\begin{document}
\vspace*{-1cm}
\begin{flushleft}
{\normalsize \rm
DESY 03-155, SFB/CPP-03-47\\
hep-ph/0310181\\
October 15, 2003, rev. December 29, 2003
}
\end{flushleft}

\vspace*{0.5cm}

\begin{center}

{\Large \bf Isospin violating effects in $\epm$ vs. $\tau$
measurements of the pion form factor $|F_\pi|^2(s)$.%
\footnote{Work supported
in part by TMR, EC-Contract No.~HPRN-CT-2002-00311 (EURIDICE), and by
DFG under Contract SFB/TR 9-03.} }

\vspace*{5mm}

{\sc Stephane Ghozzi\footnote{On leave from \'{E}cole Normale
Sup\'{e}rieure, Departement de Physique, Paris} 
and Fred Jegerlehner\footnote{\tt e-mails: stephane.ghozzi@ens.fr, fred.jegerlehner@desy.de}}

\vspace*{3mm}

{\normalsize \it
Deutsches Elektronen Synchrotron \\
Platanenallee 6, D-15738 Zeuthen, Germany}

\par
\end{center}
\vskip 5mm
\begin{center}
\bf Abstract
\end{center}
{\it We study possible so far unaccounted isospin breaking effects in
the relation between the pion form factor as determined in $e^+e^-$
experiments and the corresponding quantity obtained after accounting for
known isospin breaking effects by an isospin rotation from the
$\tau$--decay spectra. In fact the observed 10\% discrepancy in the
respective pion form factors may be explained by the isospin breaking
which is due to the difference between masses and widths of the charged
and neutral $\rho$ mesons. Since the hadronic contribution to the muon
anomalous magnetic moment can be calculated directly in terms of the
$\epm$--data the corresponding evaluation seems to be more reliable. Our
estimate is $\amu^{\mathrm{had}(1)}=(694.8 \pm
8.6)\:\times\:10^{-10}$. The $\tau$--data are useful at the presently
aimed level of accuracy only after appropriate input from theory.}
\par
\vskip 6mm

\renewcommand{\thefootnote}{\arabic{footnote}}
\setcounter{footnote}{0}

\section{Introduction}

The most precise measurement of the low energy pion form factor in
$\epm$--annihilation experiments is from the CMD-2 collaboration.  The
updated results for the process $e^+e^- \to \rho \to \pi^+\pi^-$ have
just been published~\cite{CMD2}. The update appeared necessary due to
an overestimate of the integrated luminosity in previous analyses. The
latter was published in 2002~\cite{CMD}. A more progressive error
estimate (improving on radiative corrections, in particular) allowed a
reduction of the systematic error from 1.4\% to 0.6 \% .

Since 1997 precise $\tau$--spectral functions became
available~\cite{ALEPH,OPAL,CLEO} which, to the extent that flavor
$SU(2)_{\rm f}$ in the light hadron sector is a symmetry, allows to
obtain the iso--vector part of the $\epm$--cross
section~\cite{tsai}. In this way $\tau$ data may help to substantially
improve our knowledge of $|F_\pi|^2(s)$, which is an important input
for the evaluations of the hadronic vacuum polarization contributions
to the anomalous magnetic moment of the muon $\amu$ and of the
effective fine structure constant $\aemz$ an important input for
LEP/SLC precision physics (see e.g.~\cite{EJ95}).  The idea to use the
$\tau$ spectral data to improve the evaluation of the hadronic
contributions $\amuh$ and $\dalh$ was pioneered in~\cite{ADH98}.

With increasing precision of the low energy data it more and more
turned out that we are confronted with a serious obstacle to further
progress: in the region just above the $\omega$--resonance, the
isospin rotated $\tau$--data, corrected for the known isospin
violating effects~\cite{CEN}, do not agree with the $\epm$--data at
the 10\% level~\cite{DEHZ}. Before the origin of this discrepancy is
found it will be hard to make progress in pinning further down
theoretical uncertainties in the predictions for $\amu$ and $\aemz$.

In this context isospin breaking effects in the relationship between
the $\tau$-- and the $\epm$--data have been extensively investigated
in~\cite{CEN}. One point which in our opinion has not been
satisfactorily clarified is the role of the isospin breaking effects
in the charged vs. neutral $\rho$ line--shape, which must manifest
themselves in $\dmrho$ and $\dgrho$.  Looking at the particle data
tables~\cite{PDG02}, there is no established non-zero result as
yet. Earlier statements about the problem in~\cite{ADH98,CEN,DEHZ}
adopted essentially the PDG estimate $\dmrho=0\pm 1 \ \mv$. There are
theoretical arguments about why this mass difference is expected to be
very small: the usual argument assumes $\Delta m^2_\rho \simeq \Delta
m^2_\pi$ via a sum rule, which then yields $m_{\rho^-}-m_{\rho^0}\simeq\ha
\frac{\Delta m^2_\pi}{m_{\rho^0}}\sim 0.014~
\mv ~!$ This is based, however, on an {\em assumption} which 
need not be true for the mass definition adopted in recent $\rho$--line
shape analyses\footnote{For a more detailed theoretical estimate
see~\cite{Bijnens:1996kg} and references therein (see
also~\cite{Feuillat:2000ch,Melikhov:2003hs}). Mass and width of an unstable particle,
and in particular of the $\rho$, depend on the precise definition. In
that sense they are pseudo-observables which depend on theoretical
input. What we need here is a consistent prescription to extract them
from the experimental data ($\rho$--line shape) and make sure that we
compare comparable quantities (with respect to the handling of vacuum
polarization effects, final state radiation, energy dependence of width,
background etc.).}. If fact a more recent analysis of the CMD-2 (before
the last update) and the ALEPH and CLEO data yielded $\dmrho=2.6 \pm 0.8
\ \mv$ and $\dgrho=3.1 \pm 1.7 \ \mv $~\cite{Davier:2003te} where the
uncertainty is our estimate.  The corresponding isospin corrections
may still look too small to account fully for the observed discrepancy
in the spectral functions but they clearly point towards a substantial
reduction of the problem. Our strategy therefore here is a different
one. Our hypothesis is that as a leading effect the discrepancy very
likely is due to the isospin breaking by the charged vs. neutral
$\rho$--meson parameters. A similar but subleading contribution is
expected to come from possible isospin violations in the respective
backgrounds (encoded usually by the $\rho',\rho''$ contributions). Since
the fit formulae adopted, like the Gounaris-Sakurai formula~\cite{GS68},
are far from being based on first principles we should not trust to much
in the fitting procedures based on them. E.g., usually just a set of
resonances $\rho,\omega,\rho',\rho''$ is included but we have no idea
about the background (continuum) which also should be included
somehow. We also would like to advocate that, as the level of accuracy
of the
present discussion advances, in future one should compile charged and neutral
$\rho$ data separately.

Whether the observed discrepancy is an experimental problem, or just a
so far underestimated isospin breaking effect will also be settled,
hopefully, by new results for hadronic $\epm$ cross--sections which
are under way from KLOE, BABAR and BELLE. These experiments, running
at fixed energies, are able to perform measurements via the radiative
return method~\cite{KLOE,Solodov:2002xu,Rodrigo:2001kf}. Results
presented recently by KLOE seem to agree very well with the final
CMD-2 $\epm$--data.

\section{The $\tau$ vs. $\epm$ problem}

The iso-vector part of $\sigma(e^+e^- \to {\rm hadrons})$ may be
calculated by an isospin rotation from $\tau$--decay spectra, to the
extent that the so--called conserved vector current (CVC) would be
really conserved (which it is not, see below). The relation may be
derived by comparing the relevant lowest order diagrams
which for the $\epm$ case translates into
\begin{equation}
\sigma^{(0)}_{\pi\pi}\equiv 
\sigma_0 (e^+e^- \to \pi^+\pi^-) = \frac{4\pi\alpha^2}{s}\: v_0(s)
\label{eepp}
\end{equation}
and for the $\tau$ case into
\begin{equation}
\frac{1}{\Gamma} \frac{d\Gamma}{ds}
 (\tau^- \to \pi^-\pi^0 \nu_\tau) = 
\frac {6\pi |V_{ud}|^2 S_{EW}}{m_\tau^2}
\frac{B(\tau^-\rightarrow \nu_\tau\,e^-\,\bar{\nu}_e)}
{B(\tau^-\rightarrow \nu_\tau\,\pi^- \pi^0)}\: 
\left(1-\frac{s}{m_\tau^2}\right)
\left(1 + \frac {2s}{m_\tau^2}\right)\: v_-(s)
\label{taupp}
\end{equation}
where $|V_{ud}|=0.9752\pm0.0007$~\cite{PDG02} denotes the CKM weak
mixing matrix element and $S_{\mathrm{EW(new)}}=1.0233\pm0.0006$
[$S_{\mathrm{EW(old)}}=1.0194$] accounts for electroweak radiative
corrections~\cite{Marciano:vm,Braaten:1990ef,Decker:1994ea,Erler:2002mv,CEN,DEHZ}.
The \sfs\ are obtained from the corresponding invariant mass
distributions. The $B(i)$'s are branching ratios. SU(2) symmetry (CVC)
would imply
\begin{equation}
v_-(s) =  v_0(s)\;\;.
\label{CVCrel}
\end{equation}
The spectral functions $v_i(s)$ are related to the pion form factors
$F^i_\pi(s)$ by
\begin{equation}
v_i(s)=\frac{\beta_i^3(s)}{12\pi} |F^i_\pi(s)|^2\;\;;\;\;\;(i=0,-)
\label{sfvsff}
\end{equation}
where $\beta_i^3(s)$ is the pion velocity. The difference in phase
space of the pion pairs gives rise to the relative factor
$\beta^3_{\pi^-\pi^0}/\beta^3_{\pi^-\pi^+}$~\cite{ADH98,Czyz:2000wh}.

It is important to check what precisely the experimental data in each
case represent. In CMD-2 $\epm$ measurements we exclude final state
radiation (FSR) as well as vacuum polarization effects\footnote{Tab.~1
of~\cite{CMD2} lists $|F_\pi|^2$, which includes VP but not FSR, as
well as $\sigma^0_{\pi\pi(\gamma)}$ which includes FSR but not VP. VP
and FSR are separately known quantities which we may add and subtract
according to our needs.} which are not included in the $\tau$ data in
first place. FSR as far as included by the measurement (virtual and
soft real radiation) has been subtracted together with the initial
state photon radiation. Hard FSR photons were rejected to a large
extent. In our analysis $F_\pi(s)$ obtained from CMD-2 is the
undressed (from VP and FSR) pion form--factor (see
e.g.~\cite{Hoefer:2001mx}). The available $\tau$ decay spectra all
include photon radiation (no subtractions were made), which hence has
to be subtracted a posteriori (the correction factor $G_\mathrm{EM}$
below), while photon vacuum polarization effects play no role (i.e.,
are not included). This is because the $\tau$ decay as a charged
current (CC) process proceeds by the heavy $W$ exchange, which makes
it an effective four--fermion interaction with Fermi constant $G_F$ as
a coupling in place of $\alpha(s)$. In contrast to $\alpha$ the Fermi
coupling $G_F$ is not running up to LEP energy scales. Electroweak
short distance corrections (hadronic relative to leptonic channel)
give rise to the correction factor $S_{\rm EW}=1+\delta_{\rm EW}$,
which is dominated by a leading large logarithm $(1+(\alpha/\pi)\:\ln
(M_Z/m_\tau))$ which should be resummed using the renormalization
group~\cite{Marciano:vm}. Note that the overall coupling drops out
from the ratios in (\ref{taupp}). This also makes it evident that the
subtraction of the large and strongly energy dependent vacuum
polarization effects (see e.g. Fig.~1 in~\cite{FJ03}) necessary for
the $\epm$--data, which seems to worsen the $\epm$ vs. $\tau$ problem,
was properly treated in previous analyses.

\vspace*{3.3cm}

\begin{figure}[ht]
\begin{picture}(120,60)(5,0)
\includegraphics[scale=0.6]{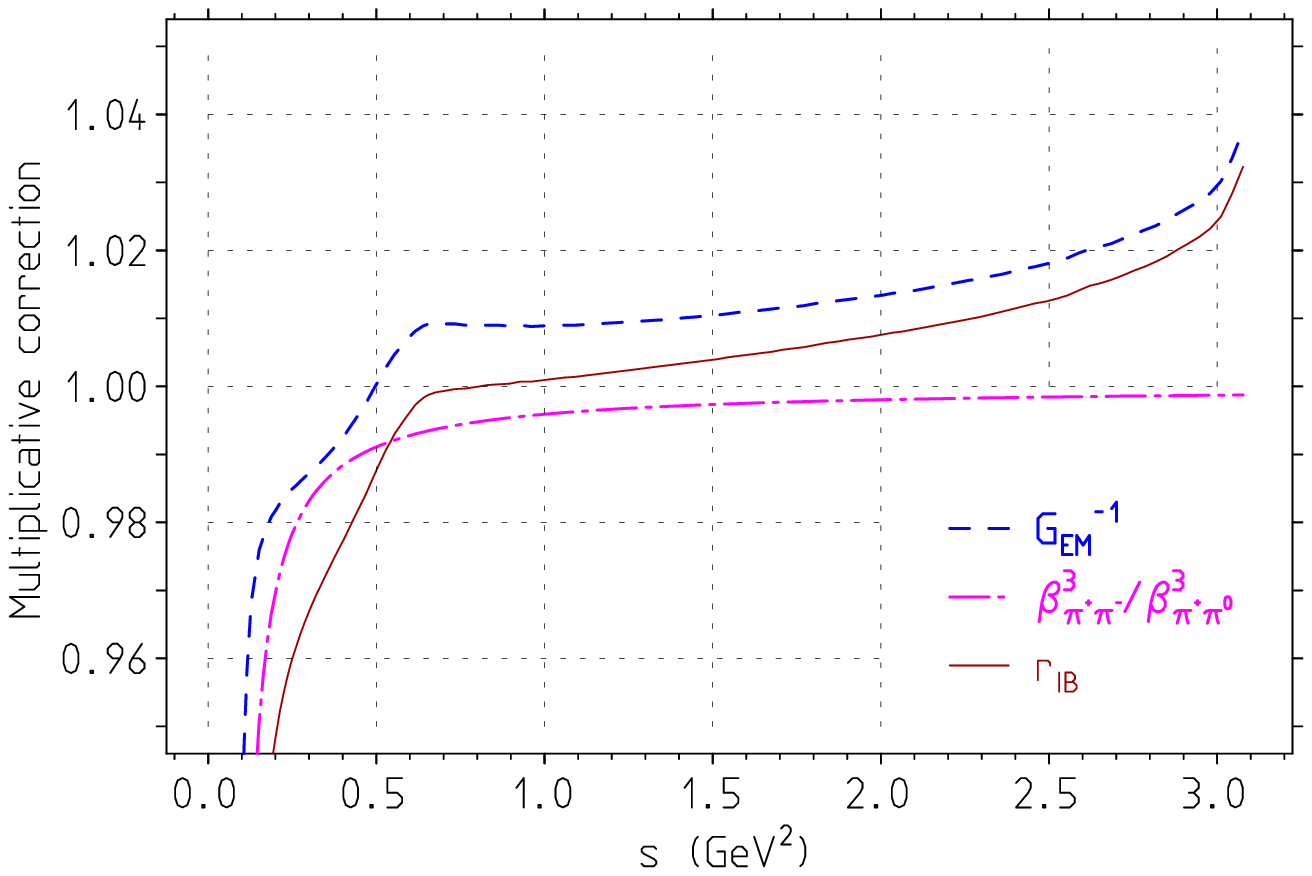}
\end{picture}

\vspace*{-2.150cm}

\begin{picture}(120,60)(-230,0)
\includegraphics[scale=0.6]{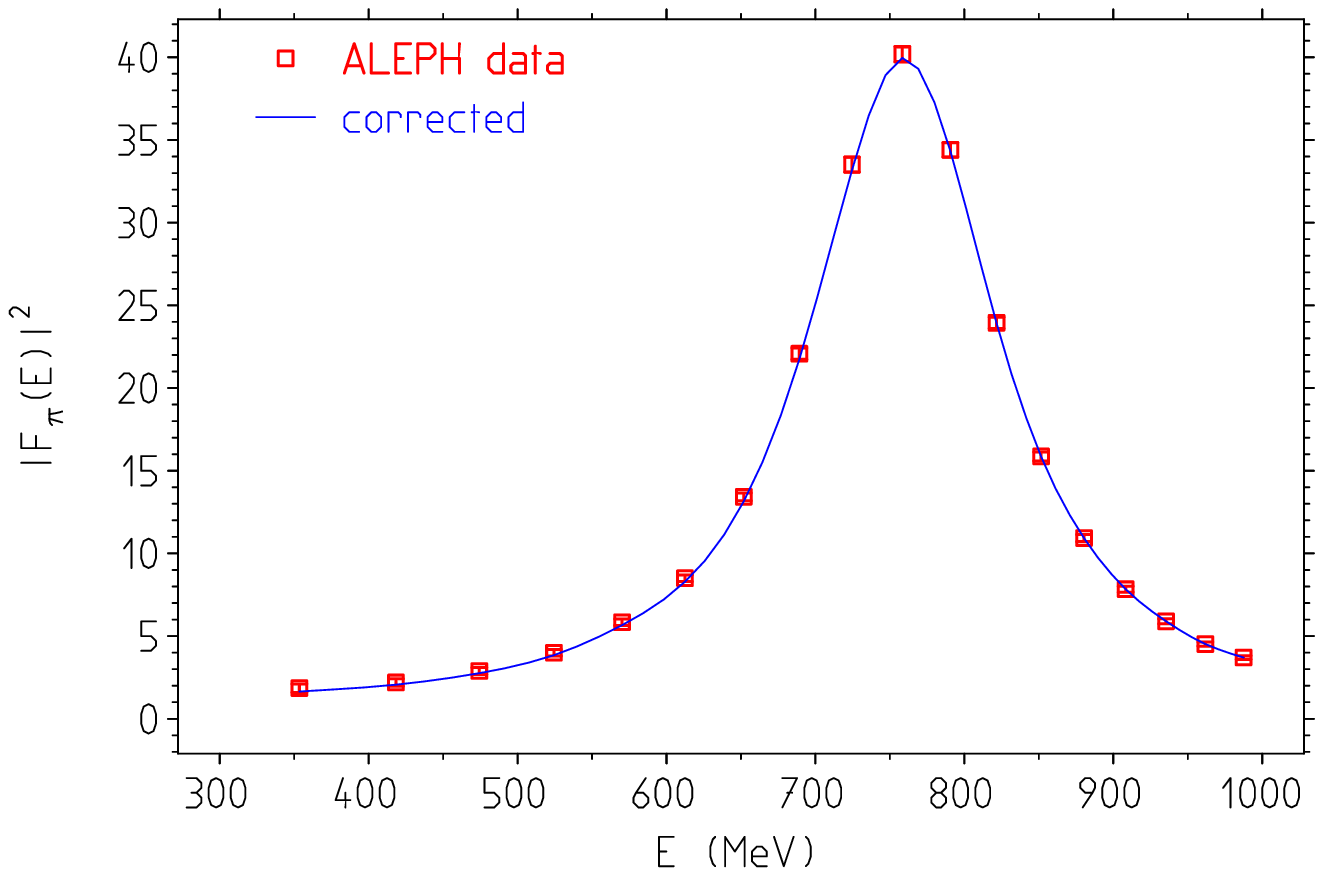}
\end{picture}


\caption[]{Isospin corrections applied to the $\tau$ data: left the
corrections~\cite{CEN}, with $r_{IB}$ defined in (\ref{rib}), and
right the barely visible effect on the ALEPH data.}
\label{fig:taudatcor}
\end{figure}

\begin{figure}[t]
\begin{picture}(120,60)(5,82)
\includegraphics[scale=0.6]{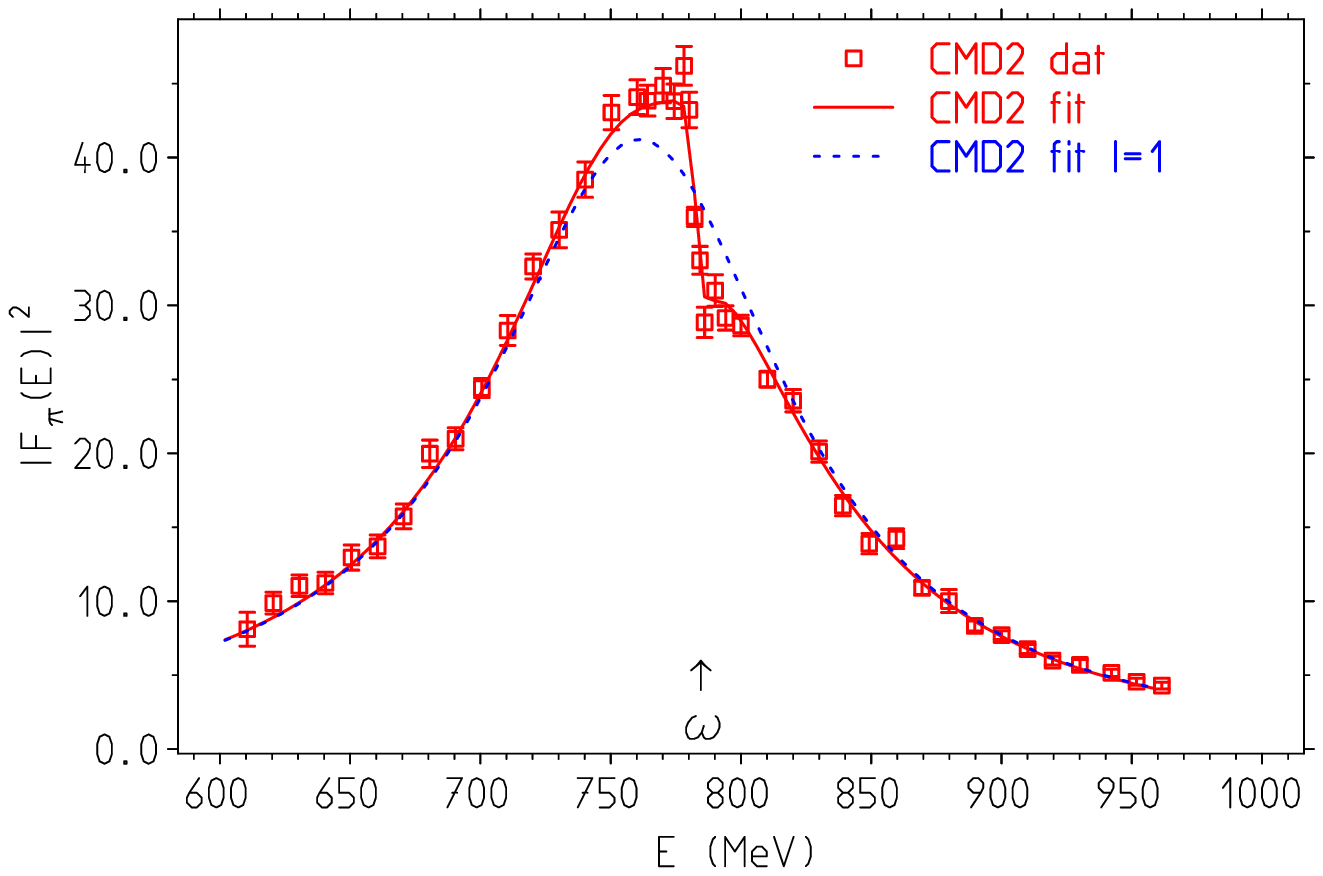}
\end{picture}

\vspace*{-2.15cm}

\begin{picture}(120,60)(-230,82)
\includegraphics[scale=0.6]{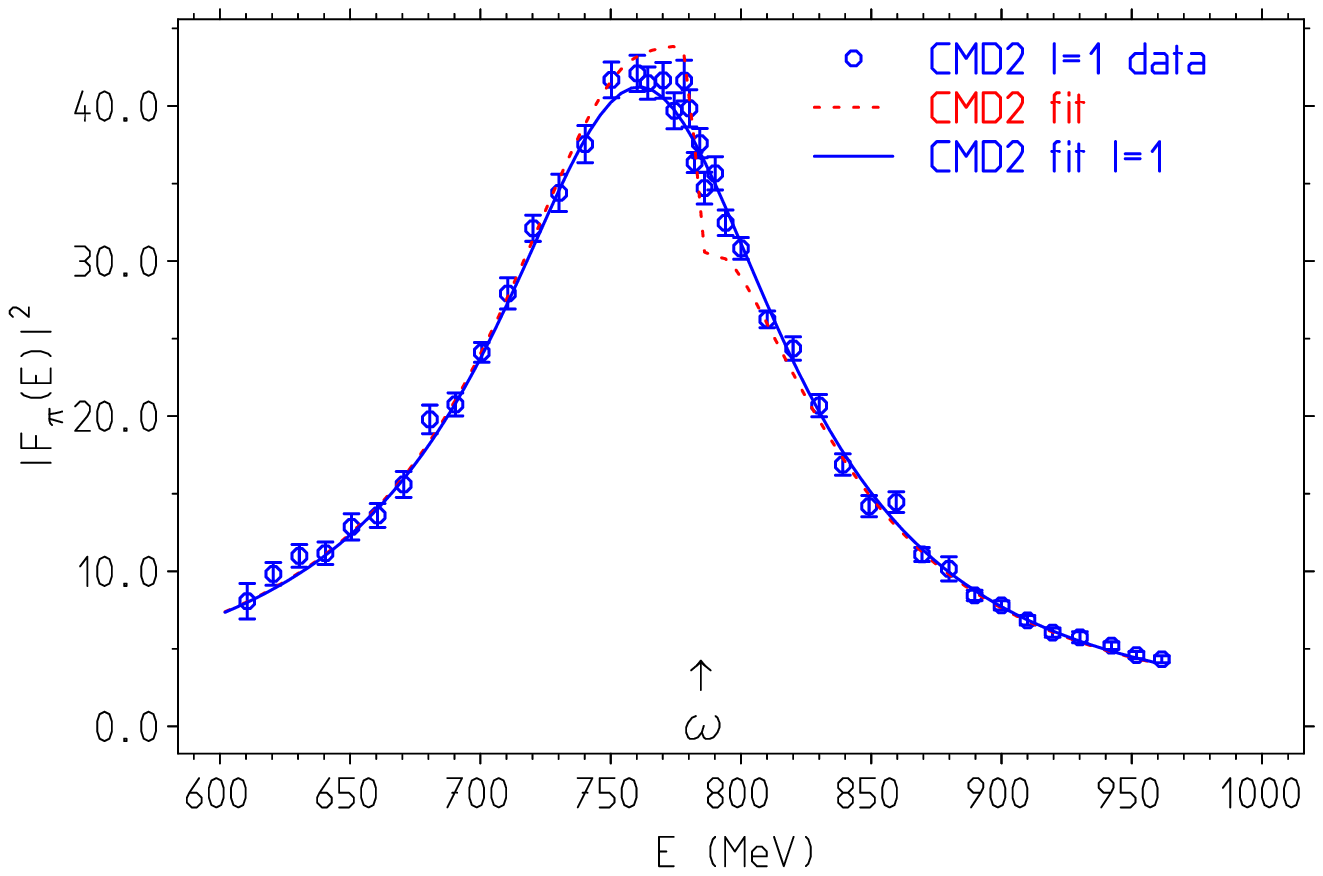}
\end{picture}

\vspace*{2.5cm}

\caption[]{CMD-2 data for $|F_\pi|^2$ in $\rho-\omega$ region 
together with Gounaris-Sakurai fit. Left: before subtraction, right:
after subtraction of the $\omega$.}
\label{fig:rhoomegasubtr}
\end{figure}

Before a precise comparison via (\ref{CVCrel}) is possible all kind of
isospin breaking effects have to be taken into account. As mentioned
earlier, this has been investigated carefully in~\cite{CEN} for the
most relevant $\pi \pi$ channel. Accordingly, we may write the
corrected version of (\ref{CVCrel}) (see~\cite{CEN} for details) in
the form
\ba
\sigma_{\pi\pi}^{(0)}=\left[ \frac{K_{\sigma}(s)}{K_\Gamma(s)}\right]
\: \frac{d \Gamma_{\pi\pi [ \gamma ] }}{ds} \times
\frac{R_{\rm IB}(s)}{S_{\rm EW}}
\ea
with\\[-8mm]
\ba
K_\Gamma (s) = \frac{G_F^2\:|V_{ud}|^2\:m_\tau^3 }{384 \pi^3}\;
\left(1-\frac{s}{m_\tau^2}
\right)^2\; \left( 1+2\: \frac{s}{m_\tau^2}\right)\;\;;\;\;\; K_\sigma
(s) = \frac{\pi \alpha^2}{3 s}\;,
\label{Ksigma}
\ea
and the isospin breaking correction
\ba
R_{\rm IB}(s) = \frac{1}{G_{\rm EM}(s)} \:
\frac{\beta^3_{\pi^-\pi^+}}{\beta^3_{\pi^- \pi^0}} \: 
\left| \frac{F_V(s)}{f_+(s)}\right|^2 \;\;.
\ea
The factor $G_{\rm EM}(s)$, displayed in Fig.~\ref{fig:taudatcor},
includes the QED corrections to $\tau^- \to \nu_\tau \pi^-
\pi^0$ decay with virtual plus real soft and hard (integrated over
all phase space) photon radiation calculated in scalar QED, except
from the short distance term, which is calculated for the
corresponding quark production process and included conventionally in
$S_{\rm EW}$ as the leading logarithm, mentioned before. Originating
from (\ref{sfvsff}), $\beta^3_{\pi^-\pi^+}/\beta^3_{\pi^- \pi^0}$ (see
Fig.~\ref{fig:taudatcor}) is a phase space correction due to the
$\pi^\pm - \pi^0$ mass difference.  $F_V(s) = F^0_\pi(s)$ is the NC
vector current form factor, which exhibits besides the $I=1$ part an
$I=0$ contribution. The latter $\rho - \omega$ mixing term is due to
the SU(2) breaking ($m_d-m_u$ mass difference). Finally,
$f_+(s)=F^-_\pi$ is the CC $I=1$ vector form factor. One of the
leading isospin breaking effects is the $\rho - \omega$ mixing
correction included in $|F_V(s)|^2$. The transition $|F_V(s)|^2 \to
|F^{(I=1)}_V(s)|^2 \sim |f_+(s)|^2$ is illustrated in
Fig.~\ref{fig:rhoomegasubtr}. The form--factor corrections, in
principle, also should include the electromagnetic shifts in the
masses and the widths of the $\rho$'s. Up to this last mentioned
effect, which was considered to be small, all the corrections were
applied in~\cite{DEHZ} but were not able to eliminate the observed
discrepancy between $v_-(s)$ and $v_0(s)$ (see ~\cite{DEHZ} for
details and Fig.~\ref{fig:ratios} below).

In fact, possible isospin breaking corrections due to different
electromagnetic shifts of masses and widths of the neutral and charged
$\rho$--mesons (remember that for other quark bound states the $\pi$'s
we have $m_{\pi^\pm}-m_{\pi^0}=4.5935\pm0.0005 \ \mv $), respectively,
have been mentioned or were very briefly discussed only
in~\cite{ADH98,CEN,DEHZ}, and they might have been underestimated so
far. Such isospin violating mass and width differences are not
established experimentally, not even the sign. We therefore ask the
question whether applying an isospin breaking correction which
accounts for that could resolve the puzzle of the above mentioned
discrepancy.

What we do is the following: we take the CMD-2 data and subtract off the
VP and the $\omega$--contribution. The latter $I=0$ part enters via
$\rho-\omega$ mixing, which is a consequence of isospin violation due
to the mass difference $m_u-m_d$ of the light quarks. To this end we may
take the Gounaris--Sakurai kind parameterization (see~\cite{GS68,CMD}
for details) (FSR not included; BW=Breit-Wigner; BW$^{\rm
GS}$=Gounaris-Sakurai modified Breit-Wigner )
\be
F_\pi(s)=\frac{\mathrm{BW}^\mathrm{GS}_{\rho(770)}(s)\cdot
\left(1+\delta \frac{s}{M_\omega^2}\mathrm{BW}_{\omega}(s)
\right)+\beta\: \mathrm{BW}^\mathrm{GS}_{\rho(1450)}(s)
     +\gamma\: \mathrm{BW}^\mathrm{GS}_{\rho(1700)}(s)}{1+\beta+\gamma}
\label{GSform}
\ee 
of the CMD-2 data and set the mixing parameter $\delta=0$. In this way
we obtain the iso--vector part of the square of the pion form factor
$|F_\pi|^{2\:{I=1}}_{(\epm)}(s)$ displayed in
Fig.~\ref{fig:rhoomegasubtr}. To the $\tau$ version of the pion form
factor, following from (\ref{taupp}) and (\ref{sfvsff}), we perform
the isospin breaking corrections
\be
r_{\mathrm{IB}}(s)=\frac{1}{G_{\rm EM}(s)} \:
\frac{\beta^3_{\pi^-\pi^+}}{\beta^3_{\pi^- \pi^0}} 
\:\frac{S_{\mathrm{EW(old)}}}{S_{\mathrm{EW(new)}}}
\label{rib}
\ee
with $G_{\rm EM}(s)$ from~\cite{CEN}. The factor
$S_{\mathrm{EW(old)}}/S_{\mathrm{EW(new)}}$ corrects for some
previously missing corrections~\cite{Erler:2002mv,DEHZ}. The such
obtained corrected\footnote{The velocity factor correction of course
only applies when, as frequently has been done, the wrong velocity was
used in (\ref{sfvsff}) in the extraction of the charged channel form
factor.}  pion form factor $|F_\pi|^{2\:{I=1}}_{(\tau)}(s)$ is to be
compared with $|F_\pi|^{2\:{I=1}}_{(\epm)}(s)$. The ratio shows the
unexpected large deviations from unity (see
Fig.~\ref{fig:ratios}). While the ALEPH and CLEO data clearly exhibit
the structure as expected from an increase of mass and width of the
$\rho$ the OPAL data show a different form of the spectrum. The
problem with the OPAL data originates from the fact that in the
neighborhood of the $\rho$ peak the cross section apparently is too
low. Since the distribution is normalized to the very precisely known
total branching fraction, the tails of the resonance get enhanced,
which leads to the structure actually seen (the apparent width gets
enhanced). Of course the data points of the spectrum have imposed
strong error correlations via the normalization to the integral
rate. In the figure only the diagonal elements of the covariance
matrix are visualized.


\begin{figure}[t]
\begin{picture}(120,60)(5,82)
\includegraphics[scale=0.6]{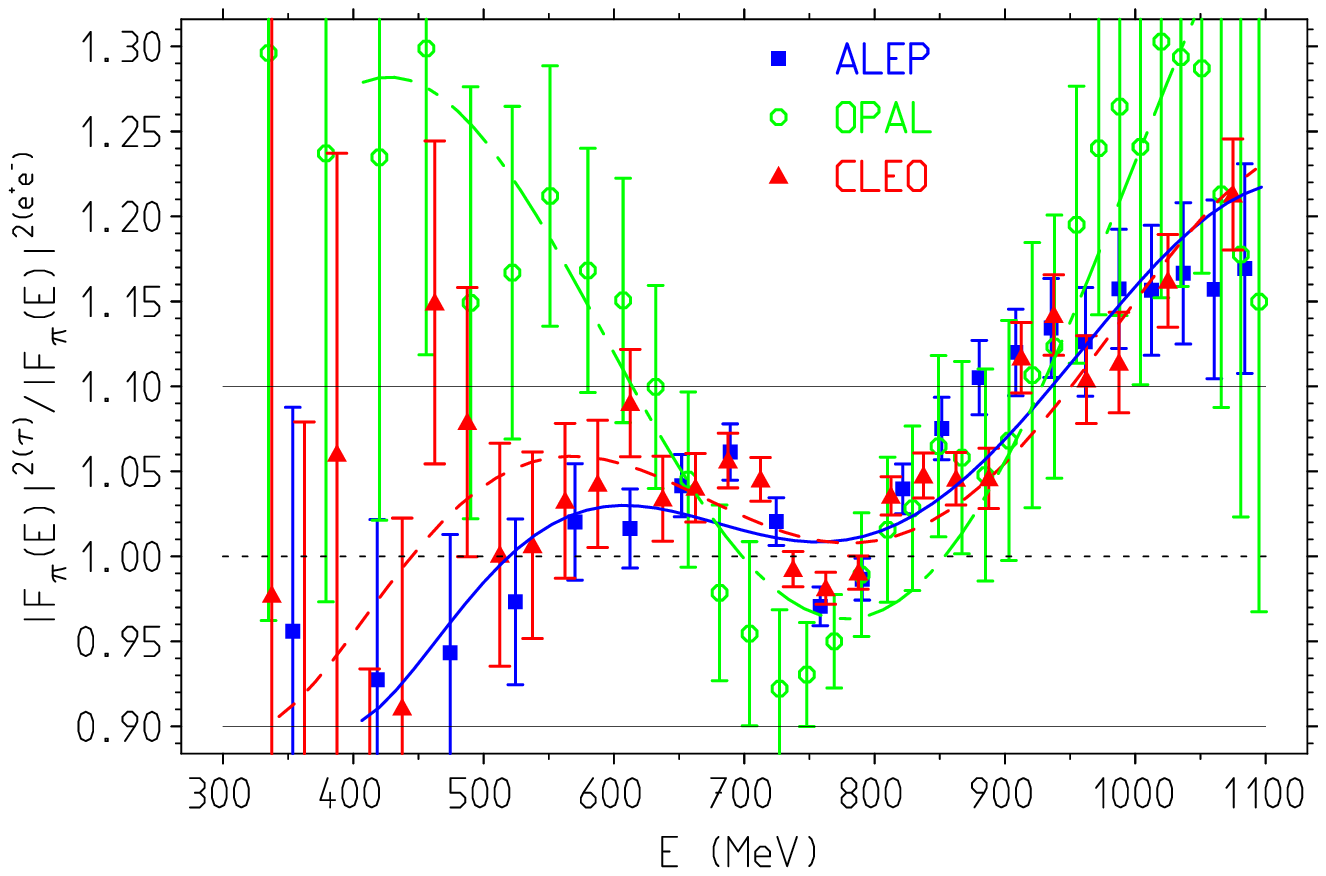}
\end{picture}

\vspace*{-2.15cm}

\begin{picture}(120,60)(-230,82)
\includegraphics[scale=0.6]{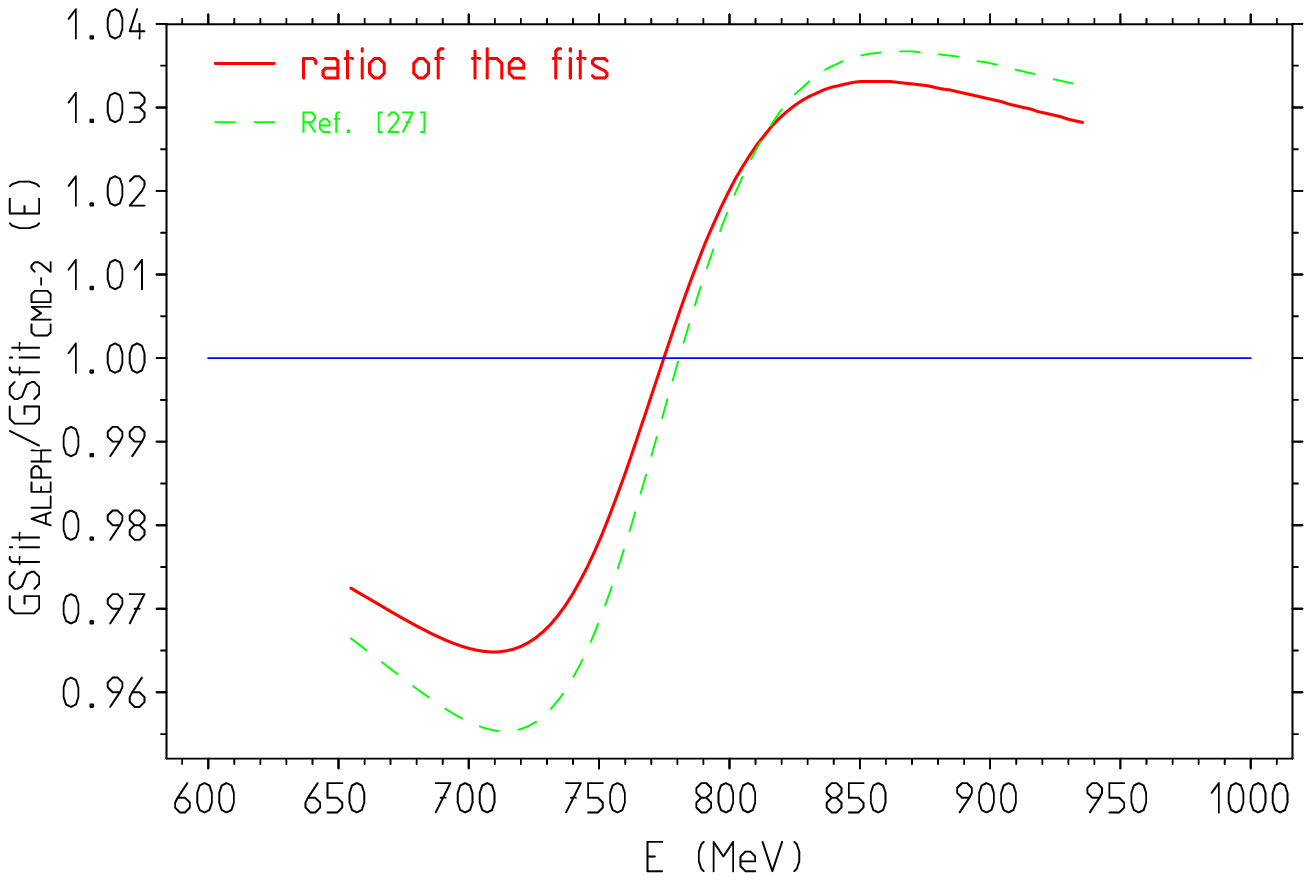}
\end{picture}

\vspace*{2.6cm}

\caption[]{Left: The ratio between $\tau$--data sets from ALEPH, OPAL and CLEO 
and the $I=1$ part of the CMD-2 fit of the $\epm$--data. The curves
which should guide the eye are fits of the ratios using 8th order
Tschebycheff polynomials. Right: Ratio of the ALEPH vs. CMD-2 fits
differing by mass and width only (see Tab.~\ref{tab:fits}). By the
isospin violation correction $(m_{\rho^-},\Gamma_{\rho^-}) \to
(m_{\rho^0},\Gamma_{\rho^0})$ of the $\tau$--data this ratio becomes
trivially equal to unity.}
\label{fig:ratios}
\end{figure}


For a comparison of the $\epm$ with the $\tau$ data we have to perform
fits in a common energy range. It is important to fix the
``$\rho$--background'' appropriately. The latter is represented in the
GS--parametrization by the higher resonances $\rho^{'}$ and
$\rho^{''}$ and in order to fit the corresponding parameters we have
to extend the fit range towards higher energies. To this end we are
including beyond the CMD-2 limit the available $\pi^+\pi^-$ data from
other experiments up to 1.5 GeV above which the $\tau$ data get of low
quality. We finally decided to adopt the PDG values for $m_{\rho'}=1465
\pm 25,~\Gamma_{\rho'}=310\pm 60$ and $m_{\rho''}=1700 \pm
20,~\Gamma_{\rho''}=240 \pm 60$ for the masses and widths and fit the
complex admixing coefficients $\beta$ and $\gamma$. For
$|\beta|=0.12\pm 0.005$, $\varphi_\beta=160\pm 5$ and $|\gamma|=0.023\pm
0.005$, $\varphi_\gamma=0$ we get reasonable fits for all data sets. In
particular the widths depend substantially on the precise values of
these parameters. Fortunately, the parameter shifts of interest turn
out to be rather stable. In other words, while there are different
possible parametrizations of comparable quality for the individual
data sets the systematic shift between $\epm$ and $\tau$ data
is there independently of how we parametrize the data.

We may fit now $|F_\pi|^{2\:{I=1}}_{(\tau)}(s)$ with the
Gounaris--Sakurai formula (\ref{GSform}) with no $\omega$ term, i.e.,
with $\delta=0$, in order to obtain $m_{\rho^\pm}$ and
$\Gamma_{\rho^\pm}$. We would like to emphasize that it is important to
``zoom--in the $\rho$'' appropriately in determining $\dmrho$ and
$\dgrho$, i.e., we have to perform the fits at fixed background (i.e.,
besides $m_\rho$ and $\Gamma_\rho$ all other parameters in the
GS-formula are held fixed) in the $\rho$ dominated region between 610.5
MeV and 961.5 MeV (CMD-2 range).

This simple leading effect analysis yields the results in
Tab.~\ref{tab:fits}\footnote{For the $\epm$ channel we fit the data
after subtraction of the $I=0$ part. Including the $\omega$ in the fit
yields practically the same result.}.  As can be seen the $\tau$ data
give consistently larger values for both mass and width of the charged
$\rho$. The evidence is far from impressive, between ALEPH and CMD-2
we have $\Delta m_\rho = 2.6 \pm 0.8 $ MeV and $\Delta \Gamma_\rho =
1.5
\pm 1.0$ MeV. The two--parameter fits are not of good quality and a more
elaborate analysis would be needed to come to more precise
conclusions. An analysis based on more recent preliminary ALEPH data,
yields the slightly larger values $\Delta m_\rho = 3.1 \pm 0.9$ MeV and
$\Delta \Gamma_\rho = 2.3
\pm 1.6$ MeV~\cite{DavierPISA}, which are consistent however with our
findings.

\begin{table*}
\begin{center}
{
\begin{tabular}{|c|c|c|c|c|} \hline 
           &  $\tau$ ALEPH & $\tau$ CLEO & $\tau$ OPAL & $e^+e^-$ CMD-2   \\
\hline \hline
$m_{\rho^0}$      & - & - & - 
&772.95 $\pm$ 0.56 $\pm$ 0.12  \\
$\Gamma_{\rho^0}$ & - & - & - 
& 147.93 $\pm$ 0.70$^{+0.12}_{-0.13}$  \\
$m_{\rho^-}$      & 775.52 $\pm$ 0.49$^{+0.34}_{-0.25}$ 
& 775.01 $\pm$ 0.36$^{+0.30}_{-0.29}$
& 777.34 $\pm$ 1.21$^{+0.29}_{-0.23}$ 
& -  \\
$\Gamma_{\rho^-}$ & 149.40 $\pm$ 0.68$^{+0.10}_{-0.09}$
& 149.00 $\pm$ 0.49$^{+0.12}_{-0.21}$
& 153.91 $\pm$ 1.62$^{+1.15}_{-1.50}$
& - \\
$S$ & 1.39 & 1.35 & 0.62 & 1.28 \\
\hline \hline
\end{tabular}
}\caption{ Results of fits to the isospin breaking corrected pion
form factors squared for $\tau$ (ALEPH, CLEO and OPAL) and $\epm$ (CMD-2)
data. The Gounaris-Sakurai parameterization of the $\rho$ line shape
is utilized. Masses and widths in MeV. $S=\sqrt{\chi^2/(n-1)}$}
\label{tab:fits}
\end{center}
\end{table*}

Note that our values of $m_{\rho^0}$ and $\Gamma_{\rho^0}$ differ
substantially from the values $775.65\pm 0.64\pm 0.5$ and $143.85\pm
1.33\pm 0.80$, respectively, given by CMD-2. The latter have been
obtained by fitting the physical $\pi\pi$ cross-section (before
subtracting the VP) and including the $\omega$ and the $\rho'$ with
mixing parameters $|\delta|=(1.57 \pm 0.16) \times 10^{-3}$,
$\varphi_\delta = 13.3^\circ \pm 3.7^\circ$ and $|\beta|=0.0695\pm
0.0053$, $\varphi_\beta=180^\circ$. A $\rho''$ was not included.  The
parameters determined by CMD-2, cannot be directly compared with
corresponding parameters obtained by fits of the $\tau$ data, because
the latter do not include VP effects. In order to have a common basis
we subtract the VP from the $\epm$ data (of course, alternatively, we
could supplement the $\tau$ data with the VP [by replacing $\alpha$ in
$K_\sigma$ (\ref{Ksigma}) by $\alpha(s)$] in order to obtain dressed
(physical) parameters which are the ones usually listed in the
particle data tables). The subtraction of the VP lowers the mass by
about 1 MeV and increases the width by about 1.3 MeV.  The additional
changes are due to the inclusion of the $\rho''$ and the changes in
the other ``background parameters''. If we would utilize the CMD-2
background parameters as a common background parametrization we
would not be able to get acceptably good fits for the $\tau$ data
sets.

Now we assume that the systematic deviations seen in the $\rho^\pm$
parameters include electromagnetic isospin breaking which we have to
correct for.  We now may ask two questions. The first is: how does the
test--ratio of Fig.~\ref{fig:ratios} look like if we replace
$m_{\rho^\pm}$ and $\Gamma_{\rho^\pm}$ in the $\tau$--data fit by the
more appropriate $m_{\rho^0}$ and $\Gamma_{\rho^0}$ ? The second is:
what mass and width do we get if we fit them in the $\tau$--data
parameterization such that the test--ratio comes out to be unity within
errors?  Not too surprisingly we find them close to the ones given in
Tab.~\ref{tab:fits} for CMD-2: It makes the central value of the ratio
unity within 0.1 \% !  Uncertainties may be obtained from the ones in
the parametrizations. Of course keeping the background fixed the result
looks pretty trivial. In fact fitting all parameters of the GS formula
simultaneously in the much wider range of $\tau$ data, as has been
performed in~\cite{Davier:2003te}, yields results which look very
similar to ours. The parameters obtained are very strongly correlated
and all of them may be affected by some isospin breaking effects. A
much more elaborate analysis would be necessary to actually establish
tight experimental values for possible isospin breakings in these
parameters.

We conclude that the $\tau$ vs. $\epm$ discrepancy very likely is an
isospin breaking effect which has not been accounted for correctly
in previous analyses. This also would establish a significant
difference for $m_{\rho^\pm}-m_{\rho^0}$ and
$\Gamma_{\rho^\pm}-\Gamma_{\rho^0}$.  Of course, what it means is that
the $\tau$ data cannot be utilized to calculate $\amuh$ without
reference to the $\epm$ data. Also, since now substantially
correlated, the inclusion of the $\tau$--data is much less
straightforward. The question is how much they still can contribute to
reduce the uncertainties in the evaluation of $\amuh$. This also makes
it very likely that the $\epm$--data based evaluations are the more
trustworthy ones. After the correction in the normalization of the
CMD-2 data we get the leading hadronic contribution to the anomalous
magnetic moment of the muon. We now obtain
\be
\amu^{\mathrm{had}(1)}=(694.8 \pm 8.6)\:\times\:10^{-10} 
~~~~[\epm-\mathrm{data \ based}].
\ee
With this estimate we get
\be
\amu^\mathrm{the} = (11\:659\:179.4 \pm 8.6_{\rm had } \pm 3.5_{\rm LBL} \pm
0.4_{\mathrm{QED+EW}}) \times 10^{-10}
\ee
which compares to the most recent experimental result~\cite{BNL}
\be
\amu^\mathrm{exp} = (11\:659\:203 \pm 8) \times 10^{-10}\;\;.
\ee
The ``discrepancy'' $|\amu^\mathrm{the} - \amu^\mathrm{exp}| = (23.6
\pm 12.3) \times 10^{-10}$ corresponds to a deviation of about 1.9
$\sigma$. For other recent estimates we refer to~\cite{DEHZ,HMNT}. 

\section{Summary and Conclusion}
Since recently we have in each case two reasonably consistent sets of
data: the ALEPH and CLEO $\tau$--data sets on the one hand and the CMD-2
and KLOE (still preliminary) $\epm$--data sets on the other hand.  The
$\tau$--data samples are about 10\% higher than the $\epm$ ones in the
tail above the $\rho$. This can be clearly seen in
Fig.~\ref{fig:ratios}. Assuming that the experiments are essentially
correct we think that a 10\% increase in a resonance tail can easily be
attributed to a 0.5\% increase in the energy scale.  Since it is very
unlikely a problem of energy calibration, the only explanation remains
that the resonance parameters must be different in the charged and the
neutral channel. Our estimated shifts in the $\rho$ parameters account
for about 6.8\% (or 8.1\% with the estimates given in~\cite{DavierPISA})
in the cross section (see Fig.~\ref{fig:ratios}). The remaining
''discrepancy'' is likely due to corresponding shifts in the other GS
fit--parameters.

Our analysis shows that to a large extent we may understand the $\epm$
vs. $\tau$ discrepancy as an isospin breaking effect coming from the
fact that mass and width of charged and neutral $\rho$--mesons, as
naively expected, are different and thus that the $\tau$--data must be
mapped to the neutral channel parameters before they can be utilized
for the evaluation of $\amuh$ in addition to the $\epm$--data. We thus
assume that a main part of the problem is due to additional isospin
breaking effects and not primarily an experimental one. Of course
there are also experimental difficulties which hopefully will be
resolved by forthcoming experiments.

The fact that the $\rho$--mass difference is found to be of size
comparable (maybe half of it) to the well--established pion mass
difference $m_{\pi^\pm}-m_{\pi^0} \sim 4.6 \ \mv $ seems not so
unlikely because the corresponding bound states, apart from the spin
orientation, have the same quark content ($\rho^\pm$ vs. $\pi^\pm$ on
the one hand and $\rho^0$ vs. $\pi^0$ on the other
hand)\footnote{Remember that, for example, para- (spin $s=0$) and
ortho- ($s=1$) positronium have almost the same binding energy. For
the pion and the $\rho$ the average distance of the quarks is
determined by the strong interactions, and in principle could be
different. However, the charge radius of the $\rho$ is of similar size
as the one of the pion and thus also the electromagnetic effects are
expected to be of similar magnitude.}. In the charged case the
electric potential is repulsive while in the neutral case it is
attractive and contributes to lower the mass, irrespective of the spin
orientation of the constituents. We do not think that the
Goldstone-boson nature of the pions, which derives from the properties
of the strong interactions, the spontaneous breaking of the chiral
symmetry, necessarily makes such a manifest isospin breaking effects
completely different for the vector particles and the scalars.

Since we do not have an independent evaluation
of the charged and neutral $\rho$--meson parameters, the isospin
correction needed in order for the $\tau$--data to be useful for the
evaluation of the hadronic contribution to the muon anomalous magnetic
moment cannot be performed at sufficient precision at the
moment. Nevertheless, the $\tau$--data still provide important cross
checks and last but not least Ref.~~\cite{ADH98} triggered a
discussion which forced all parties to check more carefully what they
have done. One impact was that also the results in the $\epm$--channel
had to be corrected.

The main point is that there must be such effects which were not
correctly treated so far. Maybe the available data do not allow us to
pin down a solid value for the mass and width differences (all GS
parameters in fact must be subject to isospin breaking and the data
may not suffice to come to a definite conclusion). We think the main
point is that the derivatives of $a_\mu$ with respect to $m_\rho$ and
$\Gamma_\rho$ are rather large and hence no stable result can be given
if the isospin breaking in these parameters is not known with
sufficient precision.

Note that in spite of the fact that the dominating $\rho$--peak is
shifted downwards, due to the correction which we have to apply to the
$\tau$--data, the $s^{-2}$ weighted $\amu$--integral does not
increase. It rather decreases, because the width also substantially
decreases by the correction and actually over-compensates the effect
of the shift in the mass.

Our conclusion: very likely we are back to \underline{one} prediction
for $a_\mu$ which is the $\epm$--based value at about $2\sigma$ below
the experimental result! Unfortunately, at present, we do not have a
precise enough understanding of the isospin violations to be able to
utilize the $\tau$--data for the evaluation of the hadronic
contribution to $g-2$ of the muon.\\

{\bf Acknowledgments\\~}

It is a pleasure to thank H.~Leutwyler, A.~Stahl, K.~M\"onig and
S.~Eidelman for numerous fruitful discussions.\\

\end{document}